\documentclass[conference, a4paper, 10pt]{IEEEtran}
\usepackage[english]{babel}
\usepackage{amssymb,amsmath,amscd,amsfonts,amsthm,bbm,bm}
\usepackage{amsmath}
\usepackage{graphicx}
\usepackage{psfrag}
\usepackage{paralist}
\usepackage{color}
\usepackage{float}
\usepackage{relsize}
\usepackage{euscript}
\usepackage{setspace}
\usepackage{flushend}
\usepackage{tikz}
\usetikzlibrary{plotmarks}
\usepackage{pgfplots}
\usetikzlibrary{arrows,shapes,decorations,backgrounds}
\usepackage{tkz-berge}
\usepackage{calc}
\usepackage{url}
\pgfplotsset {compat = 1.16}
\usepackage[nolist,printonlyused]{acronym}      
\usepackage{cite}
\usepackage{comment}
\usepackage{ragged2e}
\usepackage[font=small,labelfont=bf]{caption}
\usepackage{subcaption}

\usepackage[left=1.62cm,right=1.62cm,top=1.9cm]{geometry}

\def\BibTeX{{\rm B\kern-.05em{\sc i\kern-.025em b}\kern-.08em
	T\kern-.1667em\lower.7ex\hbox{E}\kern-.125emX}}

\DeclareMathOperator{\diag}{diag}

\newcommand{\C}{\mathbb{C}}
\newcommand{\E}{\mathbb{E}}

\def\bx{{\bf x}}\def\bx{{\bf x}}
\def\bX{{\bf X}}
\def\bR{{\bf R}}
\def\bA{{\bf A}}
\def\bB{{\bf B}}
\def\bD{{\bf D}}
\def\bI{{\bf I}}

\def\bY{{\bf Y}}
\def\bH{{\bf H}}

\def\bV{{\bf V}}

\def\bZ{{\bf Z}}

\def\ba{{\bf a}}

\def\bw{{\bf w}}

\def\bZ{{\bf Z}}
\def\bI{{\bf I}}

\def\bv{{\bf v}}

\def\bx{{\bf x}}
\def\by{{\bf y}}
\def\bz{{\bf z}}
\def\bZ{{\bf Z}}
\def\bY{{\bf Y}}

\def\btheta{{\boldsymbol \theta}}
\def\bphi{{\boldsymbol \phi}}
\def\bnu{{\boldsymbol \nu}}

\def\bSigma{{\boldsymbol \Sigma}}

\newtheorem{assumption}{Assumption}

\begin{document}
\title{On Target Detection in the Presence of Clutter in Joint Communication and Sensing Cellular Networks}
\author{
	Julia Vinogradova$^{\star}$,
	G\'{a}bor Fodor$^{\star \star \ddag}$, 
	\\
	\small $^{\star}$Ericsson Research, Finland, Email: \texttt{julia.vinogradova@ericsson.com}\\
	\small $^{\star \star}$Ericsson Research, Sweden, E-mail: \texttt{gabor.fodor@ericsson.com} \\
	\small $^\ddag$KTH Royal Institute of Technology, Sweden. E-mail: \texttt{gaborf@kth.se}
    \thanks{The Authors were partially supported by the TUDOR (Towards Ubiquitous 3D Open Resilient Network) project, funded by UK government, and the B5GPOS (Beyond 5G Positioning) project, funded by VINNOVA, Sweden, Grant No. 2022-01640.}
	\\
}
\maketitle

\pagestyle{empty}


\begin{acronym}
	\acro{2G}{Second Generation}
	\acro{3G}{3$^\text{rd}$~Generation}
	\acro{3GPP}{3$^\text{rd}$~Generation Partnership Project}
	\acro{4G}{4$^\text{th}$~generation}
	\acro{5G}{5$^\text{th}$~generation}
	\acro{AA}{Antenna Array}
	\acro{AC}{Admission Control}
	\acro{AD}{Attack-Decay}
	\acro{ADSL}{Asymmetric Digital Subscriber Line}
	\acro{AHW}{Alternate Hop-and-Wait}
    \acro{AIC}{Akaike Information Criterion}
	\acro{AMC}{Adaptive Modulation and Coding}
	\acro{AoA}{angle of arrival}
	\acro{AoD}{angle of departure}
	\acro{AP}{Access Point}
	\acro{APA}{Adaptive Power Allocation}
	\acro{AR}{autoregressive}
	\acro{ARMA}{Autoregressive Moving Average}
	\acro{ATES}{Adaptive Throughput-based Efficiency-Satisfaction Trade-Off}
	\acro{AWGN}{additive white Gaussian noise}
	\acro{BB}{Branch and Bound}
	\acro{BD}{Block Diagonalization}
	\acro{BER}{bit error rate}
	\acro{BF}{Best Fit}
	\acro{BLER}{BLock Error Rate}
	\acro{BPC}{Binary power control}
	\acro{BPSK}{binary phase-shift keying}
	\acro{BPA}{Best \ac{PDPR} Algorithm}
	\acro{BRA}{Balanced Random Allocation}
	\acro{BS}{base station}
	\acro{CAP}{Combinatorial Allocation Problem}
	\acro{CAPEX}{Capital Expenditure}
	\acro{CBF}{Coordinated Beamforming}
	\acro{CBR}{Constant Bit Rate}
	\acro{CBS}{Class Based Scheduling}
	\acro{CC}{Congestion Control}
	\acro{CDF}{Cumulative Distribution Function}
	\acro{CDMA}{Code-Division Multiple Access}
	\acro{CL}{Closed Loop}
	\acro{CLPC}{Closed Loop Power Control}
    \acro{CM}{concurrent mode}
	\acro{CNR}{Channel-to-Noise Ratio}
	\acro{CPA}{Cellular Protection Algorithm}
	\acro{CPICH}{Common Pilot Channel}
	\acro{CoMP}{Coordinated Multi-Point}
	\acro{CQI}{Channel Quality Indicator}
	\acro{CRLB}{Cram\'er-Rao Lower Bound}
	\acro{CRM}{Constrained Rate Maximization}
	\acro{CRN}{Cognitive Radio Network}
	\acro{CS}{Coordinated Scheduling}
	\acro{CSI}{channel state information}
	\acro{CSIR}{channel state information at the receiver}
	\acro{CSIT}{channel state information at the transmitter}
	\acro{CUE}{cellular user equipment}
	\acro{D2D}{device-to-device}
	\acro{DCA}{Dynamic Channel Allocation}
	\acro{DE}{Differential Evolution}
	\acro{DFT}{Discrete Fourier Transform}
	\acro{DIST}{Distance}
	\acro{DL}{downlink}
	\acro{DMA}{Double Moving Average}
	\acro{DMRS}{Demodulation Reference Signal}
	\acro{D2DM}{D2D Mode}
	\acro{DMS}{D2D Mode Selection}
	\acro{DPC}{Dirty Paper Coding}
	\acro{DRA}{Dynamic Resource Assignment}
	\acro{DSA}{Dynamic Spectrum Access}
	\acro{DSM}{Delay-based Satisfaction Maximization}
	\acro{ECC}{Electronic Communications Committee}
	\acro{EFLC}{Error Feedback Based Load Control}
	\acro{EI}{Efficiency Indicator}
	\acro{eNB}{Evolved Node B}
	\acro{EPA}{Equal Power Allocation}
	\acro{EPC}{Evolved Packet Core}
	\acro{EPS}{Evolved Packet System}
	\acro{E-UTRAN}{Evolved Universal Terrestrial Radio Access Network}
	\acro{ES}{Exhaustive Search}
	\acro{FDD}{frequency division duplexing}
	\acro{FDM}{Frequency Division Multiplexing}
	\acro{FER}{Frame Erasure Rate}
	\acro{FF}{Fast Fading}
	\acro{FSB}{Fixed Switched Beamforming}
	\acro{FST}{Fixed SNR Target}
	\acro{FTP}{File Transfer Protocol}
	\acro{GA}{Genetic Algorithm}
	\acro{GBR}{Guaranteed Bit Rate}
	\acro{GCS}{global coordinate system}
	\acro{GLR}{Gain to Leakage Ratio}
	\acro{GOS}{Generated Orthogonal Sequence}
	\acro{GPL}{GNU General Public License}
	\acro{GRP}{Grouping}
	\acro{HARQ}{Hybrid Automatic Repeat Request}
	\acro{HMS}{Harmonic Mode Selection}
	\acro{HOL}{Head Of Line}
	\acro{HSDPA}{High-Speed Downlink Packet Access}
	\acro{HSPA}{High Speed Packet Access}
	\acro{HTTP}{HyperText Transfer Protocol}
	\acro{ICMP}{Internet Control Message Protocol}
	\acro{ICI}{Intercell Interference}
	\acro{ID}{Identification}
	\acro{IETF}{Internet Engineering Task Force}
	\acro{ILP}{Integer Linear Program}
	\acro{JCAS}{joint communication and sensing}
	\acro{JRAPAP}{Joint RB Assignment and Power Allocation Problem}
	\acro{UID}{Unique Identification}
	\acro{IID}{Independent and Identically Distributed}
	\acro{IIR}{Infinite Impulse Response}
	\acro{ILP}{Integer Linear Problem}
	\acro{IMT}{International Mobile Telecommunications}
	\acro{INV}{Inverted Norm-based Grouping}
	\acro{IoT}{Internet of Things}
	\acro{IP}{Internet Protocol}
	\acro{IPv6}{Internet Protocol Version 6}
	\acro{ISD}{Inter-Site Distance}
	\acro{ISI}{Inter Symbol Interference}
	\acro{ITU}{International Telecommunication Union}
	\acro{JOAS}{Joint Opportunistic Assignment and Scheduling}
	\acro{JOS}{Joint Opportunistic Scheduling}
	\acro{JP}{Joint Processing}
	\acro{JS}{Jump-Stay}
	\acro{KF}{Kalman filter}
	\acro{KKT}{Karush-Kuhn-Tucker}
	\acro{L3}{Layer-3}
	\acro{LAC}{Link Admission Control}
	\acro{LA}{Link Adaptation}
	\acro{LC}{Load Control}
	\acro{LCS}{Local Coordinate System}
	\acro{LOS}{line-of-sight}
	\acro{LP}{Linear Programming}
	\acro{LS}{least squares}
	\acro{LTE}{Long Term Evolution}
	\acro{LTE-A}{LTE-Advanced}
	\acro{LTE-Advanced}{Long Term Evolution Advanced}
	\acro{M2M}{Machine-to-Machine}
	\acro{MAC}{Medium Access Control}
	\acro{MANET}{Mobile Ad hoc Network}
	\acro{MB}{multibeam}
	\acro{MC}{Modular Clock}
	\acro{MCS}{Modulation and Coding Scheme}
	\acro{MDB}{Measured Delay Based}
	\acro{MDI}{Minimum D2D Interference}
    \acro{MDL}{Minimum Description Length}
	\acro{MF}{Matched Filter}
	\acro{MG}{Maximum Gain}
	\acro{MH}{Multi-Hop}
	\acro{MIMO}{multiple input multiple output}
	\acro{MINLP}{Mixed Integer Nonlinear Programming}
	\acro{MIP}{Mixed Integer Programming}
	\acro{MISO}{multiple input single output}
	\acro{ML}{maximum likelihood}
	\acro{MLWDF}{Modified Largest Weighted Delay First}
	\acro{MME}{Mobility Management Entity}
	\acro{MMSE}{minimum mean squared error}
	\acro{mmWave}{millimeter wave}
	\acro{MOS}{Mean Opinion Score}
	\acro{MPF}{Multicarrier Proportional Fair}
	\acro{MRA}{Maximum Rate Allocation}
	\acro{MR}{Maximum Rate}
	\acro{MRC}{maximum ratio combining}
	\acro{MRT}{Maximum Ratio Transmission}
	\acro{MRUS}{Maximum Rate with User Satisfaction}
	\acro{MS}{mobile station}
	\acro{MSE}{mean squared error}
	\acro{MSI}{Multi-Stream Interference}
	\acro{MTC}{Machine-Type Communication}
	\acro{MTSI}{Multimedia Telephony Services over IMS}
	\acro{MTSM}{Modified Throughput-based Satisfaction Maximization}
	\acro{MU-MIMO}{multiuser multiple input multiple output}
	\acro{MU}{multi-user}
	\acro{MUSIC}{MUltiple SIgnal Classification}
	\acro{NAS}{Non-Access Stratum}
	\acro{NB}{Node B}
	\acro{NE}{Nash equilibrium}
	\acro{NCL}{Neighbor Cell List}
	\acro{NLP}{Nonlinear Programming}
	\acro{NLOS}{non line-of-sight}
	\acro{NMSE}{Normalized Mean Square Error}
	\acro{NORM}{Normalized Projection-based Grouping}
	\acro{NP}{Non-Polynomial Time}
	\acro{NRT}{Non-Real Time}
	\acro{NSPS}{National Security and Public Safety Services}
	\acro{O2I}{Outdoor to Indoor}
	\acro{OFDMA}{orthogonal frequency division multiple access}
	\acro{OFDM}{orthogonal frequency division multiplexing}
	\acro{OFPC}{Open Loop with Fractional Path Loss Compensation}
	\acro{O2I}{Outdoor-to-Indoor}
	\acro{OL}{Open Loop}
	\acro{OLPC}{Open-Loop Power Control}
	\acro{OL-PC}{Open-Loop Power Control}
	\acro{OPEX}{Operational Expenditure}
	\acro{ORB}{Orthogonal Random Beamforming}
	\acro{JO-PF}{Joint Opportunistic Proportional Fair}
	\acro{OSI}{Open Systems Interconnection}
	\acro{PAIR}{D2D Pair Gain-based Grouping}
	\acro{PAPR}{Peak-to-Average Power Ratio}
	\acro{P2P}{Peer-to-Peer}
	\acro{PC}{Power Control}
	\acro{PCI}{Physical Cell ID}
	\acro{PDF}{Probability Density Function}
	\acro{PDPR}{pilot-to-data power ratio}
	\acro{PER}{Packet Error Rate}
	\acro{PF}{Proportional Fair}
	\acro{P-GW}{Packet Data Network Gateway}
	\acro{PL}{Pathloss}
	\acro{PPR}{pilot power ratio}
	\acro{PRB}{physical resource block}
	\acro{PROJ}{Projection-based Grouping}
	\acro{ProSe}{Proximity Services}
	\acro{PS}{Packet Scheduling}
	\acro{PSAM}{pilot symbol assisted modulation}
	\acro{PSO}{Particle Swarm Optimization}
	\acro{PZF}{Projected Zero-Forcing}
	\acro{QAM}{Quadrature Amplitude Modulation}
	\acro{QoS}{Quality of Service}
	\acro{QPSK}{Quadri-Phase Shift Keying}
	\acro{RAISES}{Reallocation-based Assignment for Improved Spectral Efficiency and Satisfaction}
	\acro{RAN}{Radio Access Network}
	\acro{RA}{Resource Allocation}
	\acro{RAT}{Radio Access Technology}
	\acro{RATE}{Rate-based}
    \acro{RMT}{random matrix theory}
	\acro{RB}{resource block}
	\acro{RBG}{Resource Block Group}
	\acro{REF}{Reference Grouping}
	\acro{RF}{radio frequency}
	\acro{RLC}{Radio Link Control}
	\acro{RM}{Rate Maximization}
	\acro{RNC}{Radio Network Controller}
	\acro{RND}{Random Grouping}
    \acro{ROC}{receiver operating characteristics}
	\acro{RRA}{Radio Resource Allocation}
	\acro{RRM}{Radio Resource Management}
	\acro{RSCP}{Received Signal Code Power}
	\acro{RSRP}{Reference Signal Receive Power}
	\acro{RSRQ}{Reference Signal Receive Quality}
	\acro{RR}{Round Robin}
	\acro{RRC}{Radio Resource Control}
	\acro{RSSI}{Received Signal Strength Indicator}
	\acro{RT}{Real Time}
	\acro{RU}{Resource Unit}
	\acro{RUNE}{RUdimentary Network Emulator}
	\acro{RV}{Random Variable}
	\acro{SAC}{Session Admission Control}
	\acro{SB}{single beam}
	\acro{SCM}{Spatial Channel Model}
	\acro{SC-FDMA}{Single Carrier - Frequency Division Multiple Access}
	\acro{SD}{Soft Dropping}
	\acro{S-D}{Source-Destination}
	\acro{SDPC}{Soft Dropping Power Control}
	\acro{SDMA}{Space-Division Multiple Access}
	\acro{SE}{spectral efficiency}
	\acro{SER}{Symbol Error Rate}
	\acro{SES}{Simple Exponential Smoothing}
	\acro{S-GW}{Serving Gateway}
	\acro{SINR}{signal-to-interference-plus-noise ratio}
	\acro{SI}{Satisfaction Indicator}
	\acro{SIP}{Session Initiation Protocol}
	\acro{SISO}{single input single output}
	\acro{SIMO}{Single Input Multiple Output}
	\acro{SIR}{signal-to-interference ratio}
	\acro{SLNR}{Signal-to-Leakage-plus-Noise Ratio}
	\acro{SMA}{Simple Moving Average}
	\acro{SNR}{Signal-to-noise ratio}
    \acro{SCNR}{Signal-to-clutter-plus-noise ratio}
	\acro{SORA}{Satisfaction Oriented Resource Allocation}
	\acro{SORA-NRT}{Satisfaction-Oriented Resource Allocation for Non-Real Time Services}
	\acro{SORA-RT}{Satisfaction-Oriented Resource Allocation for Real Time Services}
	\acro{SPF}{Single-Carrier Proportional Fair}
	\acro{SRA}{Sequential Removal Algorithm}
	\acro{SRS}{Sounding Reference Signal}
	\acro{SU-MIMO}{single-user multiple input multiple output}
	\acro{SU}{Single-User}
	\acro{SVD}{Singular Value Decomposition}
	\acro{TCP}{Transmission Control Protocol}
	\acro{TDD}{time division duplexing}
    \acro{TDM}{time division mode}
	\acro{TDMA}{Time Division Multiple Access}
	\acro{TETRA}{Terrestrial Trunked Radio}
	\acro{TP}{Transmit Power}
	\acro{TPC}{Transmit Power Control}
	\acro{TTI}{Transmission Time Interval}
	\acro{TTR}{Time-To-Rendezvous}
	\acro{TSM}{Throughput-based Satisfaction Maximization}
	\acro{TU}{Typical Urban}
	\acro{UE}{user equipment}
	\acro{UEPS}{Urgency and Efficiency-based Packet Scheduling}
	\acro{UL}{uplink}
	\acro{ULA}{uniform linear array}
	\acro{UMTS}{Universal Mobile Telecommunications System}
	\acro{URI}{Uniform Resource Identifier}
	\acro{URM}{Unconstrained Rate Maximization}
	\acro{UT}{user terminal}
	\acro{VR}{Virtual Resource}
	\acro{VoIP}{Voice over IP}
	\acro{WAN}{Wireless Access Network}
	\acro{WCDMA}{Wideband Code Division Multiple Access}
	\acro{WF}{Water-filling}
	\acro{WiMAX}{Worldwide Interoperability for Microwave Access}
	\acro{WINNER}{Wireless World Initiative New Radio}
	\acro{WLAN}{Wireless Local Area Network}
	\acro{WMPF}{Weighted Multicarrier Proportional Fair}
	\acro{WPF}{Weighted Proportional Fair}
	\acro{WSN}{Wireless Sensor Network}
	\acro{WWW}{World Wide Web}
	\acro{XIXO}{(Single or Multiple) Input (Single or Multiple) Output}
	\acro{ZF}{zero-forcing}
	\acro{ZMCSCG}{Zero Mean Circularly Symmetric Complex Gaussian}
\end{acronym}

\begin{abstract}
Recent works on joint communication and sensing (JCAS) cellular networks have proposed to use  time division mode (TDM) and concurrent mode (CM), as alternative methods for sharing the resources between communication and sensing signals. While the performance of these JCAS schemes for object tracking and parameter estimation has been studied in previous works,
their performance on target detection in the presence of clutter has not been analyzed.
In this paper, we propose a detection scheme for estimating the number of targets in JCAS cellular networks that employ TDM or CM resource sharing. The proposed detection method allows for the presence of clutter and/or temporally correlated noise. This scheme is studied with respect to the JCAS trade-off parameters that allow to control the 
time slots in TDM and the power resources in CM allocated to sensing and communications.
The performance of two fundamental transmit beamforming schemes, typical for JCAS, is compared in terms of the receiver operating characteristics curves.
Our results indicate that in general
the TDM scheme gives a somewhat better detection performance compared to the CM scheme, although both schemes outperform existing approaches provided that their respective trade-off parameters are tuned properly.
\end{abstract}


\acresetall 

\section{Introduction}
The evolution of cellular networks and their enabling technologies are driven
by both the insatiable demands for mobile broadband and Internet of Things connectivity services and location-aware and sensing applications \cite{Hammarberg:22, Behravan:22}.
Indeed, highly reliable location-aware and sensing services provided by transport, smart city, automated factory and personalized health care applications are expected to become one of the attractive new features of emerging 6G networks \cite{Behravan:23, Wang:23}. Due to their integrated communication, localization and sensing capabilities, 6G systems are evolving to perceptive mobile networks \cite{Zhang:21}, which can be seen as a natural evolution of point-to-point dual-function communication and radar systems \cite{Li:16, Paul:17}.

One of the key enabling technologies of perceptive mobile networks is \ac{JCAS}, which builds on the key observation that wireless communication and radar sensing share many commonalities in terms of hardware, signal processing algorithms and system architecture \cite{Liu:20,Zhang:21b,Liu:22,Liu:22d,JAZhang:22}. Recognizing the large interest in prospective sensing services provided by the ubiquitous cellular infrastructure, there is an increasing interest in developing energy and spectral efficient \ac{JCAS} protocols and algorithms by both the academic and standardization communities, including the \ac{3GPP} and the IEEE \cite{S1-214242, Wu:22, Behravan:22}.

Conceptually, there are two distinct approaches to sharing the time-space resources between communication and sensing signals in \ac{JCAS} systems, which can be referred to as \ac{TDM} and \ac{CM}, capitalizing on widely deployed \ac{MIMO} systems and advanced multi-beam solutions \cite{Liu:20, Zhang:19, Barneto:20, Baig:23}.
As the name suggests, in \ac{TDM}, the communication and radar signals are separated in the time domain, such that at any one particular time instant, a slot or a frame is allocated to either a communication or a radar (sensing) signal, and orthogonality between the signals is maintained. In contrast, in \ac{CM}, the communication and sensing signals may use the same or overlapping time and frequency resources, since either the same signal is used for carrying information to the intended receiver and for detecting and estimating the characteristics of passive objects, or the signals are separated in the spatial or beam domains \cite{Liu:20, Zhang:19}. It is intuitively clear that both modes need to deal with their respective inherent trade-offs, which are related to the portion of time and the portion of the overall energy made available for communication and sensing \cite{Baig:23}. 

In particular, the recently proposed multibeam technology \cite{Liu:20, Zhang:19, Barneto:20} enables to form spatially distinct lobes using either hybrid or analog beamforming architectures. In particular multibeam (actually referring to creating multiple lobes by analog beamforming) using a single \ac{RF} chain and multiple transmit antennas was proposed by \cite{Zhang:19}, and further optimized by \cite{Barneto:20}. In contrast, \cite{Liu:20} uses hybrid beamforming architectures with multiple \ac{RF} chains to create multiple beams for serving communicating \acp{UE} and transmitting and receiving sensing signals concurrently.
In our recent work \cite{Baig:23}, we compared the performance of the \ac{TDM} and the \ac{CM} in a bistatic \ac{JCAS} system in terms of the achieved \ac{CRLB} for \ac{AoA} estimation of passive objects (targets) and the achievable \ac{SINR} at the served \ac{UE}. The overall conclusion of these recent works on using multibeams, as a key enabler for the \ac{CM} operation of \ac{JCAS} systems, is that \ac{CM} may be superior to \ac{TDM} provided that the overall transmit energy is properly divided between the communication and sensing beams (lobes) \cite{Baig:23}.  

Along a closely related research line, it has been pointed out by several recent works that for target detection -- which is necessary for proper parameter estimation and tracking of passive objects -- distinguishing the desired and unwanted radar echoes and suppressing clutter are crucial and highly non-trivial steps in \ac{JCAS} systems and perceptive cellular networks \cite{Zhang:21, Zhang:21b, JAZhang:22, Malanowski:22, Ksiezyk:23}. A key difficulty for clutter suppression in \ac{JCAS} systems is that the desired multipath and the radar
echoes may be caused by the same scatterers and reflectors \cite{Zhang:19}. Therefore, for the purpose of estimating the channel parameters and then subsequently forming multibeams for \ac{JCAS}, it may be necessary to beamform towards those scatterers, as it has been pointed out in \cite{Liu:20}.
Consequently, the performance of the \ac{TDM} and \ac{CM} multibeam schemes in terms of correct target detection and false alarm rates may be different, which need to be carefully analyzed. However, to the best of our knowledge, the impact of clutters on the performance of \ac{TDM} and \ac{CM} multibeam \ac{JCAS} systems has not been modelled and analyzed in the literature. 

In this paper we propose a detection scheme for estimating the number of targets in the presence of clutters in a \ac{JCAS} system that employs either \ac{TDM} or \ac{CM} operation. The proposed target detection method not only allows for the presence of clutters, but also deals with temporally correlated noise at the sensing receiver under assumption of the noise model introduced in \cite{Vinogradova:13} in the context of radar detection. The proposed detection test is based on \ac{RMT} studied in the context of blind radar detection in the presence of white and colored noise models in \cite{Cardoso:08, Bianchi:11, Nadler:11, Vinogradova:13}. The proposed scheme is studied with respect to the JCAS trade-off parameters allowing to control the power and time resources allocated to sensing and communication respectively. The performance of two fundamental transmit beamforming schemes, for which sensing and communication beams are separated in time (TDM) and in space (CM), respectively are compared in terms of the well-known \ac{ROC} curves \cite{Dokhanchi:23}. 


We show that the proposed target detection approach allows to estimate the number of targets accurately, while the existing methods fail as they are typically designed for white noise scenarios.

\par \textit{Notations:} The superscripts $(\cdot)^{\sf T}$ and $(\cdot)^{\sf H}$
denote the transpose and Hermitian transpose operations, respectively.
The notation ${\cal CN}(a,\sigma^{2})$ represents the complex circular
Gaussian distributions with mean $a\in\mathbb{C}$ and variance
$\sigma^{2}\in\mathbb{R}^{+}$. 
The vector $\boldsymbol{0}_{n}$ denotes the $n$-dimensional vector
with entries equal to zero, and $\bI_{n}$ denotes the identity matrix
of dimension $n\times n$. 

\section{System model}\label{System_mod}

\subsection{System model}
\begin{figure}[H]
	\centering 
	\includegraphics[width=8.8cm]{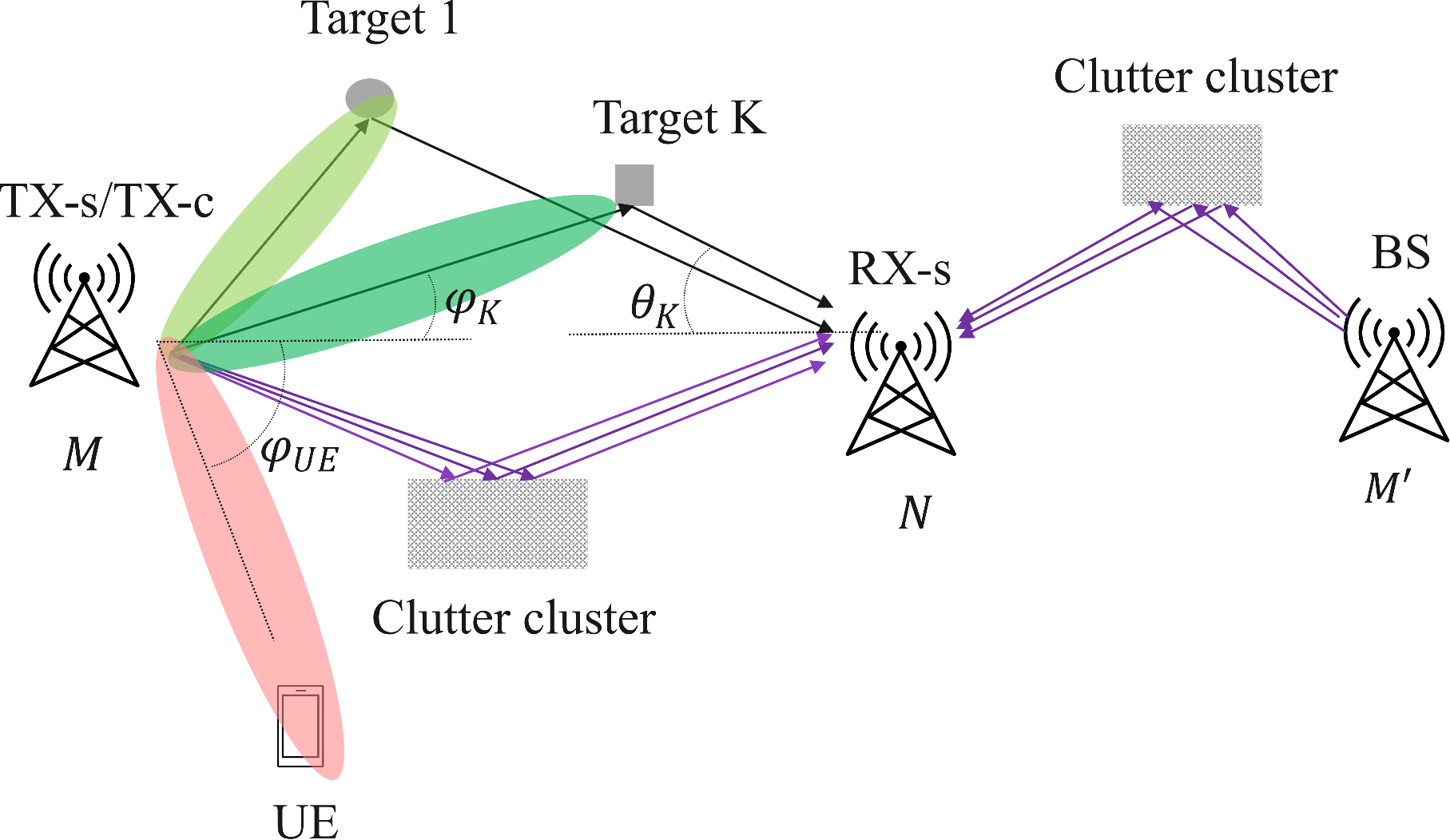}
	\caption{System model with a sensing/communication transmitting \ac{BS}, denoted by TX-s/TX-c with transmit beamforming towards $K$ targets, a served \ac{UE}, and a sensing receiver \ac{BS}, denoted by RX-s, which is used to detect the targets.}
	\label{fig:system_model}
\end{figure}

We consider a bi-static scenario with the sensing transmitter (TX) and the sensing receiver (RX) nodes located at different \acp{BS}, as illustrated in Figure~\ref{fig:system_model}. 

We assume uniform linear arrays with $M$ and $N$ antenna elements at the transmitter and the receiver sensing node, respectively, denoted by TX-s and RX-s. We assume that the transmitter is used for \ac{DL} communication with a \ac{UE}, and denote this node by TX-c during the communication phase.
We consider $K$ targets -- also referred to as passive objects since they are not equipped with wireless transmitter or receiver antennas -- modeled as single point reflectors with \ac{AoD} towards the RX-s node, denoted by $\phi_k$ for $k=1, \ldots, K$, while the \acp{AoA} at the receiver sensing node are denoted by $\theta_k$ for $k=1, \ldots, K$ (see Figure \ref{fig:system_model}).

In this system, the transmit and receive steering vectors are defined by
\begin{align}
	\ba_M(\phi)&\triangleq \frac{1}{\sqrt{M}} \left[1, e^{-j\pi \cos \phi}, \ldots, e^{-j\pi (M-1)\cos \phi} \right]^{\sf T} \in\mathbb{C}^{M\times1}  \label{steer_tx}, \\ 
 \text{~and~} \nonumber  \\
	\ba_N(\theta)&\triangleq \frac{1}{\sqrt{N}} \left[1, e^{-j\pi \cos \theta}, \ldots, e^{-j\pi (N-1)\cos \theta} \right]^{\sf T} \in\mathbb{C}^{N\times1}, \label{steer_rx}
\end{align}
respectively, where $\phi\in\left[0,2\pi\right]$ and $\theta \in\left[0,2\pi\right]$.

\subsection{Clutter model}




The \ac{NLOS} scattered signal components in millimeter wave frequencies are usually composed of clusters of points, as it has been shown e.g. in \cite{Wen:21}. Each cluster presents a set of points with a certain angular spread of the incident and scattered signals. Based on this observation, the clutter occurring in millimeter wave propagation environment can be modeled by a set of such scattered signals. 

Specifically, in this paper we assume that the (overall) clutter is composed of $L$ clusters, where cluster $l$ ($l=1,\ldots,L$) contains $J_l$ clutter points, uniformly distributed along a line. The clutter points are, in general, closely located on a scatter surface associated with a certain angular spread, which depends on the cluster's location and its physical properties. The clutter can be caused from either the same transmit antenna array as the one used for target detection or another transmit antenna array with a different number of antenna elements, as denoted by $M'$ in Figure~\ref{fig:system_model}. In the sequel, for ease of notation, we will assume that all clusters of clutters are caused by the same transmit \ac{BS} signals.

\subsection{Temporally correlated noise model}


 We consider a temporally correlated noise model similar to the model presented in \cite{Vinogradova:13}. We denote by $v_n[t]$ the noise at the receiver antenna element $n=0, \ldots, N-1$ at time instant $t=0, \ldots, T_s-1$, where $T_s$ is the total number of sensing time slots.
The auto-correlation function of the stochastic process $v_n[t]$, for $i=1-T_s, \ldots, T_s-1$ is defined as
\begin{equation} \label{r_i}
	r(i) \triangleq \E\left\{v_{n}[t]v_{n}[t+i]^*\right\}~\forall t; ~\forall n.
\end{equation}
	
The covariance matrix of $\bv_n=[v_n[0], \ldots, v_n[T_s]]^{\sf T} \in \C^{T_s \times 1}$ is given by
\begin{align}\label{Sigma}
	&\bSigma \triangleq \E \left\{ \bv_n \bv_n ^{\sf H}\right\} \nonumber \\
	&= \begin{bmatrix}
		r(0) & r(1) & \ldots & r(T_s-1) \\
		r(-1) & \ddots & \ddots & \vdots \\
		\vdots  & \ddots & \ddots & r(1)\\
		r(1-T_s) & \ldots & r(-1)  & r(0)
	\end{bmatrix} \in \C^{T_s \times T_s},
\end{align}
where $\bSigma$ is a Toeplitz matrix.

\subsection{Sensing transmission model}
In general, each transmission path component corresponding to a target or a clutter point, is associated with a given delay/range and a Doppler shift. In this paper, we do not include the Doppler and delay terms in the system model. Indeed, as the main focus of this paper is detection of the targets, including these terms results in unnecessarily additional model complexity which is not useful for analysing the proposed detection test at this initial step.

The received signal model at the sensing node at time slot $t$ 
is given by
\begin{align}
	\by\left[t\right]&=\sum_{k=1}^{K} \nu_k \ba_N(\theta_k)\ba_M^{\sf H}(\phi_k) \bx[t] \nonumber \\
	& +\sum_{l=1}^{L} \sum_{j_l=1}^{J_l} \widetilde{\nu}_{l,j_l}\ba_N(\widetilde{\theta}_{l,j_l})\ba_{M}^{\sf H}(\widetilde{\phi}_{l,j_l}) \widetilde{\bx}[t] +\bv\left[t\right]~\in\mathbb{C}^{N\times1},
	\label{eq: ReceivedSignal}
\end{align}
where $\nu_k\in\C$ denotes the path loss corresponding to the $k$-th target for $k=1, \ldots, K$, $\ba_N(\theta_k)$ and $\ba_M(\phi_k)$ are the steering vectors corresponding to the target \ac{AoD} and \ac{AoA} $\phi_k$ and $\theta_k$, respectively, and $s \left[t\right] \sim {\cal CN}(0,1)$ denotes the transmitted symbol at time instant $t$, which is assumed to be the same symbol transmitted over all the $M$ transmit antennas. The vector $\bx[t] = P_s \bw_M\left[t\right] s\left[t\right] \in\mathbb{C}^{M \times 1}$ represents the beamfomed transmitted signal vector with $\bw_M \left[t\right] \in\mathbb{C}^{M \times 1}$ denoting the transmit precoding weight vector towards the sensing direction $\phi_{s}$ at time instant $t$, and $P_s$ denotes the sensing transmission power. The precoding design for different transmission schemes will be studied in Section~\ref{tx_beamforming}.
The vectors $\ba_N(\widetilde{\theta}_{l,j_l})$ and $\ba_M^{\sf H}(\tilde{\phi}_{l,j_l})$  denote respectively the receive and transmit steering vectors corresponding to the $j_l$-th clutter point from the $l$-th cluster with \ac{AoA} $\tilde{\theta}_{l,j_l}$ and \ac{AoD} $\widetilde{\phi}_{l,j_l}$, for $l=1, \ldots, L$ and $j_l=1,\ldots, J_l$. The vector $\widetilde{\bx}[t]\in\mathbb{C}^{M \times1}$ (with complex Gaussian entries of variance $P_{\text{cl}}$) represents the transmit signal vector by a transmit antenna array causing the clutter. Finally, $\bv[t]=[v_0[t], \ldots, v_{N-1}[t]]^{\sf T} \in \C^{N \times 1}$ stands for the received noise vector at time instant $t$.

\subsection{Communication transmission model}

In this paper, we consider a simple single user \ac{MIMO} channel for the communication path and assume that the \ac{UE} is scheduled at a slot $t\in\left\{ 0,1,\dots,T_c-1\right\}$ to receive the \ac{DL} transmission, where $T_c$ is the number of communication time-slots.


Consequently, the received signal at the \ac{UE} under \ac{LOS} assumption of the communication link is
\begin{equation}
	y_{\textrm{UE}}\left[t\right]=\ba_{M}^{\sf H}(\theta_{\textrm{UE}}) \bw_{\textrm{UE}}\left[t\right] s\left[t\right]+z_{\textrm{UE}}\left[t\right]\in\mathbb{C},
	\label{eq: ReceivedSignal_com}
\end{equation}
where $\theta_{\textrm{UE}}\in\left[0,2\pi\right]$ is the \ac{AoA} of the UE, $z_{\textrm{UE}}\left[t\right]\sim\mathcal{CN}\left({0},\sigma_{\textrm{UE}}^{2}\right)$ is the white Gaussian noise and $\bw_{\textrm{UE}}\left[t\right]\in\mathbb{C}^{M\times 1}$ is the transmit precoding vector towards the \ac{UE}.

It should be noted that similar steering vector-based models for sensing in Equation~\eqref{eq: ReceivedSignal} and for communication in Equation~\eqref{eq: ReceivedSignal_com} have already been proposed in the literature \cite{Liu:20}.
We call slot $t$ a {\it communication
slot} if the \ac{UE} is scheduled to receive a \ac{DL} transmission in that slot. As we will see in the sequel, both schemes may use multiple communication slots out of the $T$ slots.

\subsection{Trade-off parameters}\label{trade_off_param}
In this paper, in order to study the trade-off between sensing and communication, we consider two main parameters $\alpha \in \left[0,1\right]$ and $\delta \in\left[0,1\right]$ allowing to control the amount of  resources in the time and power domains, respectively. The total number of time symbols that can be allocated for both sensing and communication is assumed to be equal to $T$. The total number of sensing symbols is denoted by $T_s$ and the number of communication slots is denoted by $T_c$.
The total power allocated to sensing and communication is assumed to be equal to $P$, the sensing power is denoted by $P_s$ and the communication power is denoted by $P_c$.

The \ac{TDM} and \ac{CM} resource sharing schemes and the usage of their associated trade-off parameters are discussed further in 
Section~\ref{tx_beamforming}. Both trade-off parameters are considered for the studied beamforming schemes in the performance analysis in Section~\ref{simu}.

\section{Transmit precoding for target detection}
\label{tx_beamforming}

We consider two transmission modes allowing to perform sensing and communication jointly. The first mode is the \ac{CM}, for which the sensing and communication are performed simultaneously and separated spatially. The second mode is  the \ac{TDM}, for which sensing and communication are separated in time, while the overall resource sharing is designed jointly. 

\subsection{Concurrent mode}
In the \ac{CM} transmission scheme, the transmitter employs a generalized version of the multibeam scheme \cite{Zhang:19,Baig:23} adapted for sensing of multiple targets by scanning simultaneously a target at the direction $k$, for $k=1, \ldots, K$, and forming a communication beam in the direction of the connected \ac{UE} during all $T$ time slots. The \ac{AoD} towards the \ac{UE} is denoted by $\phi_{\textrm{UE}}$, which is used to form the communication-specific transmit beam in the \ac{DL}.
The sensing transmit angles towards are denoted by $\phi_{s,k}$, for $k=1, \ldots, K$, and are used to form the sensing-specific transmit beams towards the $K$ targets. Each target direction is sensed during $T_{s,k}=\left\lfloor T/K \right \rfloor$ time slots, and $\mathcal{C}_{s,k}$ denotes the set of sensing slot indices used to sense in the direction of the target $k$.
The transmitting BS forms the transmit precoding vector in the direction of the target $k$, ($k=1, \ldots, K$), and the \ac{UE}, expressed as:
\begin{align}
\bw_{\textrm{CM}}[t] \triangleq \sqrt{\frac{\delta}{K}}\ba_M(\phi_{\text{s},k}) +\sqrt{1-\delta}~\ba_M(\phi_{\text{UE}}) \in \C^{M \times 1}
& t \in \mathcal{C}_{s,k} \label{w_tx_cm},
\end{align}
where $\ba_M(\cdot)$ is the transmit steering vector defined in Equation~\eqref{steer_tx}, $t\in \mathcal{C}_{s,k}$, and $\delta \in\left[0,1\right]$ is a parameter defined Section~\ref{trade_off_param} allowing to control the amount of power resources allocated for sensing.

\subsection{Time-division mode}
We recall that $\alpha \in \left[0,1\right]$ is a parameter allowing to control the amount of resources in the time.
In the \ac{TDM} scheme, sensing and communication are performed into two separated time phases. During the sensing phase which lasts for $T_s=\left\lfloor \alpha T \right \rfloor$ slots, the transmitter scans consecutively the direction of the target $k$ during $T_{s,k}$ time slots, which is a subset of $T_s$ slots, for $k=1,\ldots, K$. During the communication phase lasting for $T_c=T-\left\lfloor \alpha T \right \rfloor$ slots, the transmitter transmits \ac{DL} signals to the connected \ac{UE}. Hence, the \ac{TDM} transmit precoding vector at time $t$ is given by
\begin{align}
	\bw_{\textrm{TDM}} [t]& \triangleq \begin{cases}
		\frac{1}{\sqrt{K}}\ba_M(\phi_{\text{s},k}), & t \in \mathcal{C}_{s,k}\\
		\ba_M(\phi_{\text{UE}}), & t \in \mathcal{C}_c
	\end{cases},
 \label{w_tx_tdm}
\end{align}
where recall that $\mathcal{C}_{s,k}$ is the set of sensing slot indices at the direction of the target $k$, while $\mathcal{C}_{c}$ is the set of communication slot indices at the direction of the served \ac{UE}.

\subsection{Resource sharing in proposed beamforming schemes}
\label{resource_sharing}

\begin{figure}[H]
	\centering 
	\includegraphics[width=9cm]{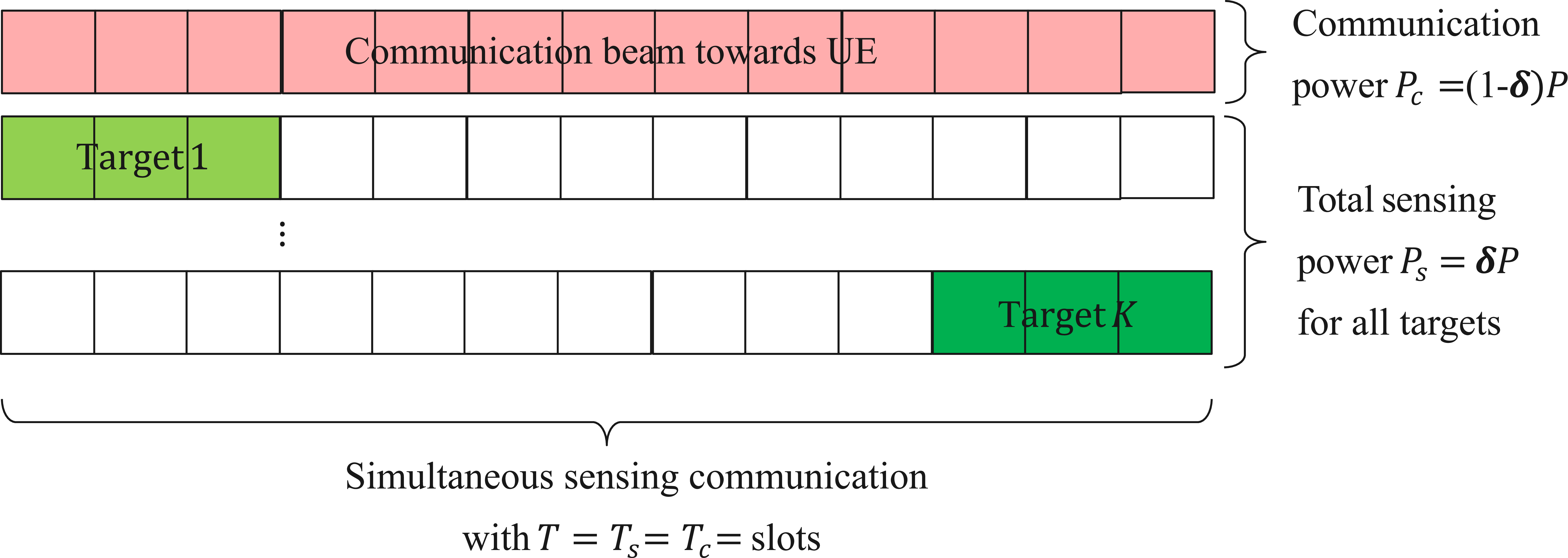}
	\caption{Space-time resource sharing for \ac{CM} beamforming scheme.
Note that each target's direction is sensed during $\left\lfloor T/K \right \rfloor$ slots.}
	\label{fig:cm}
\end{figure}

Figure~\ref{fig:cm} depicts the space-time resource sharing for the proposed \ac{CM} beamforming scheme between sensing (in green) and communication (in pale red). The total number of slots used for sensing and communication is equal to $T=T_s=T_c$ slots. The total power allocated to sensing and communication is equal to $P$. The power allocated to sensing of all the $K$ targets is denoted by $P_s=\delta P$ and $P_c=(1-\delta )P$ is the power allocated to communication with $\delta \in [0,\ldots,1]$ and $P=P_s+P_c$. Each target's direction $k$ is steered during $T_{s,k}=\left\lfloor T/K \right \rfloor$ time slots. Note that the per target power is equal to $P_s/K=\delta P/K$.
 
\begin{figure}[H]
	\centering 
	\includegraphics[width=9cm]{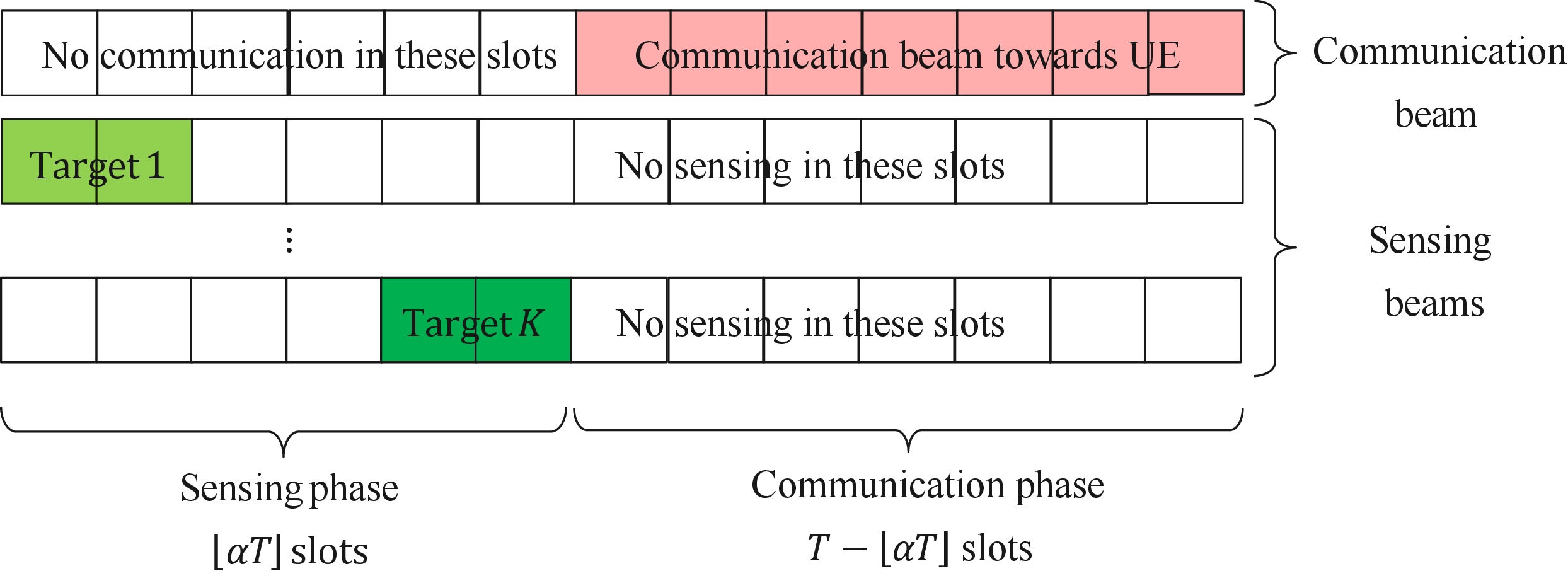}
	\caption{Space-time resource sharing for the \ac{TDM} beamforming scheme. Each target's direction is sensed during $\left\lfloor T_s/K \right \rfloor$ slots.}
	\label{fig:tdm}
\end{figure}

Figure~\ref{fig:tdm} depicts the space-time resource sharing for the proposed \ac{TDM} beamforming scheme. As indicated, the number of slots used for sensing is fixed to $T_s=\left\lfloor \alpha T\right\rfloor$ slots and the number of slots allocated for communication transmission to the \ac{UE} is equal to $T_c=T-\left\lfloor \alpha T\right\rfloor$. The total number of slots is denoted by $T=T_s+T_c$ and each target's direction $k$ is steered during $\left\lfloor T_s/K \right \rfloor$ time slots. 

Note that the parameter $\alpha$ limits the number of sensing slots and sets the communication slots available for data transmission to the \ac{UE} in the \ac{TDM} scheme. On the other hand, the parameter $\delta$ limits the energy fed into the $K$ sensing beams and the communication beam for the \ac{CM} scheme.

\section{Target detection test}

\subsection{Signal model}

The channel corresponding to the $K$ targets in Equation~\eqref{eq: ReceivedSignal}, can be written in the matrix form as
	\begin{align}\label{H}
		\bH&=\sum_{k=1}^{K}\nu_k\ba_N(\theta_k)\ba_M^{\sf H}(\phi_k)  \\ \nonumber
		&= \bA_N({\btheta}) {\bD} \bA_M^{\sf H}({\bphi}) ~~\in \C^{N \times M}
	\end{align}
with
\begin{align*}
\bA_N({\btheta}) &=[\ba_N(\theta_1), \ldots, \ba_N(\theta_K)] \in \C^{N \times K} \\ \nonumber
\bA_M({\bphi}) &=[\ba_M(\phi_1), \ldots, \ba_M(\phi_K)] \in \C^{M \times K} \\ \nonumber
\bD({\bnu}) &= \diag \left\{\nu_1, \ldots, \nu_K \right\} \in \C^{K \times K},
\end{align*}
where $\btheta=[\theta_1, \ldots, \theta_K]^{\sf T}\in \C^{K \times 1}$ is a vector containing the target \acp{AoA}, $\bphi=[\phi_1, \ldots, \phi_K]^{\sf T}\in \C^{K \times 1}$ is a vector containing the target \acp{AoD}, and  $\bnu=[\nu_1, \ldots, \nu_K]^{\sf T}\in \C^{K \times 1}$ contains the target path losses modeled as complex Gaussian variables of variance $\sigma_{{\nu}_k}^2$.

Similarly, the channel corresponding to the clutter is written as 
\begin{align}\label{H_cl}
	\bH_{\text{cl}}&=\sum_{l=1}^{L} \sum_{j_l=1}^{J_l} \widetilde{\nu}_{l,j_l} \ba_N(\widetilde{\theta}_{l,j_l})\ba_{M}^{\sf H}(\widetilde{\phi}_{l,j_l}) \\ \nonumber
		&= \bA_N(\widetilde{\btheta}){\bD}(\widetilde{\bnu}) \bA_{{M}}^{\sf H}(\widetilde{\bphi}) \C^{N \times M}
\end{align}
with
\begin{align*}
	\bA_N(\widetilde{\btheta}) &=[\ba_N(\widetilde{\theta}_1), \ldots, \ba_N(\widetilde{\theta}_{K_{\text{cl}}})] \in \C^{N \times K_{\text{cl}}}  \\ \nonumber
	\bA_{M}(\widetilde{\bphi}) &=[\ba_{M}(\widetilde{\phi}_1), \ldots, \ba_{M}(\widetilde{\phi}_{K_{\text{cl}}})] \in \C^{M \times K_{\text{cl}}} \\ \nonumber
	{\bD}(\widetilde{\bnu}) &= \diag \left\{\widetilde{\nu}_1, \ldots, \widetilde{\nu}_{K_{\text{cl}}} \right\} \in \C^{K_{\text{cl}} \times K_{\text{cl}}},
\end{align*}
where $K_{\text{cl}}=\sum_{l=1}^{L}J_l$ is the total number of clutter points, $\widetilde{\btheta}=[\widetilde{\theta}_1, \ldots, \widetilde{\theta}_{K_{\text{cl}}}]^{\sf T}\in \C^{{K_{\text{cl}}} \times 1}$ is a vector containing the clutter \acp{AoA}, 
$\widetilde{\bphi}=[\widetilde{\phi}_1, \ldots, 
\widetilde{\phi}_{K_{\text{cl}}}]^{\sf T}\in \C^{{K_{\text{cl}}} \times 1}$ 
is a vector containing the clutter  \acp{AoD}, and  $\widetilde{\bnu}=[\widetilde{\nu}_1, \ldots, \widetilde{\nu}_{K_{\text{cl}}}]^{\sf T}\in \C^{{K_{\text{cl}}} \times 1}$ contains the clutter path losses modeled as complex Gaussian variables of variance assumed to take different values for different clusters and all equal for the clutter points from a given cluster $\sigma_{\widetilde{\nu}_l}^2$, with $l=1,\ldots, L$.

Concatenating the received noise vector over the observation window of size $T_s$, the noise matrix is given by	$\bV=[\bv[0], \ldots, \bv[T_s-1]]\in \C^{N \times T_s}$ which, due to its temporal correlation property, can be equivalently written as a product of two matrices \cite{Vinogradova:13}:
\begin{equation}\label{V_mtx}
		\bV= \bZ \bSigma^{1/2} \in \C^{N \times T_s},
\end{equation}
where $\bZ=[\bz[0], \ldots, \bz[T_s-1]] \in \C^{N \times T_s}$ is the white noise matrix with complex Gaussian i.i.d. entries of variance $\sigma^2$ and $\bSigma^{1/2}$ is a square root of the noise covariance matrix $\bSigma\in \C^{N \times N}$ defined in Equation~\eqref{Sigma}. 
	
We concatenate $T_s$ independent observations of the received signal defined by the model in Equation~\eqref{eq: ReceivedSignal} and obtain the received observation matrix $\bY=[\by[0], \ldots, \by[T_s-1]] \in \C^{N \times T_s}$. The corresponding received signal model can be written in the matrix form, using Equations~\eqref{H}, \eqref{H_cl} and \eqref{V_mtx}, as 
\begin{equation} \label{rx_matrix_mod}
     \bY = \bH\bX+ \bH_{\text{cl}}\widetilde{\bX} + \bV, 
\end{equation}
where $\bX= [\bx[0], \ldots, \bx[T_s-1]] \in \C^{M \times T_s}$ is the transmitted signal matrix containing concatenated beamformed transmitted signal vectors towards the $K$ targets (with transmit beamforming vectors as defined in Equations~\eqref{w_tx_cm} and~\eqref{w_tx_tdm}) over all the sensing time slots and $\widetilde{\bX}=[\widetilde{\bx}[0], \ldots, \widetilde{\bx}[T_s-1]]\in \C^{M \times T_s}$ is the matrix of transmitted signals contributing to the clutter. Both $\bX$ and $\widetilde{\bX}$ are assumed to be unitary matrices.

We notice that the received signal matrix defined in Equation~\eqref{rx_matrix_mod} is written as a sum of three matrices corresponding to the targets, the clutter, and the noise.
Hence, we define the following matrices:
\begin{align}
        \bB_{\text{tar}} &\triangleq \bH \bX  \label{B_tar}\\
 		\bB_{\text{cl}} &\triangleq \bH_{\text{cl}} \widetilde{\bX}
   \label{B_cl},
\end{align}
where $\bB_{\text{tar}}$ represents the target-related part of the received signal and $\bB_{\text{cl}}$ corresponds to the clutter part. 

To proceed, we define the sample covariance matrix \cite{Couillet:11} based on the $T_s$ observations of the received signal which can be written using \eqref{rx_matrix_mod} as
\begin{equation}\label{cov_mtx}
    {\bR}_{T_s} \triangleq \frac{1}{T_s} \bY\bY^{\sf H} \in \C^{N \times N}.
\end{equation}


The main result of this paper is based  on the usage of the eigenvalues of the matrix ${\bR}_{T_s}$.
In general, as $\bY$ is composed from matrices $\bB_{\text{tar}}$, $\bB_{\text{cl}}$, and $\bV$, which all contain random variables, the \ac{RMT} tools can be used in order to characterize the limit spectrum of ${\bR}_{T_s}$. We discuss the assumptions on the system parameters in the following section that will allow us to apply the results from \ac{RMT} to our system model in Equation~\eqref{rx_matrix_mod}.

\subsection{Fixed-rank perturbation models}

\par The emergence of the massive \ac{MIMO} technology \cite{Asplund:20} makes usage of larger antenna arrays as compared to the previous generation wireless systems. Hence, the size of the received signal vector used for signal processing methods applied at the receiver side is getting increasingly large. In classical wireless systems the size of the received signal vector is small as compared to the number of its available time samples. In massive \ac{MIMO}, it is realistic to assume that the size of the transmitted or received signal by a \ac{BS} antenna array is of the same order of magnitude as the number of available time samples.
The number of targets is assumed to be small compared to the number of receiver antennas. 

Hence, we assume in the following that $M$, $N$, and $T_s$ are large and of the same order of magnitude. $K$ is assumed to be fixed as $M$, $N$, $T_s$ grow large. We assume the absolute summability of the noise covariance coefficients defined in Equation~\eqref{r_i} resulting in a bounded sum $\sum_{i=0}^{T_s-1}|r(i)| \leq C$ where $C$ is a positive fixed constant as $N, T_s \to \infty$. These assumptions are resumed into the following:

\begin{assumption}
	As $N \to \infty$, $T_s \to \infty$, $N/T_s \to {c}>0$.
\end{assumption}

\begin{assumption}
	As $N\to \infty$, $M \to \infty$, $M/N \to \bar{c}>0$.
\end{assumption}

\begin{assumption}
	$K$ is fixed as $N, T \to \infty$.
\end{assumption}

\begin{assumption}
	$\sum_{i=0}^{N-1}|r(i)| \leq C$, $C>0$ as $N, T_s \to \infty$.
\end{assumption}

Under the above assumptions, the matrix form system model given by Equation~\eqref{rx_matrix_mod} corresponds to the fixed rank perturbation model in \ac{RMT} \cite{Baik:06}, \cite{Hachem:13}. It should be noted that the assumption “large” is useful for applying the \ac{RMT} results to our system model. In practice, as we will see in the simulation analysis, $M$, $N$, and $T_s$ take realistic values corresponding to typical system parameters in the currently deployed wireless communication systems.

\subsubsection{Clutter-free case}
In order to understand the main idea, we consider first the clutter-free case for which the received model, written as $\bB_{\text{tar}}+\bV$, corresponds to a sum of two random matrices, with $\bB_{\text{tar}}$ defined by Equation~\eqref{B_tar}. The random matrix corresponding to the temporally correlated noise $\bV=\bZ\bSigma^{1/2}$ is of full rank equal to $\text{min}(N,T_s)$ with probability one. Under Assumptions~1 and 4, it is known that the spectrum of its sample covariance matrix $\frac{1}{T_s}\bV\bV^{\sf H}$ converges to a compactly supported probability measure with probability one \cite{Chapon:14}, \cite{Marcenko:67}. This means that the eigenvalues of $\frac{1}{T_s}\bV\bV^{\sf H}$ form one ``bulk'' of eigenvalues with probability one.

The random matrix corresponding to the targets $\bB_{\text{tar}}$ is of fixed rank equal to $K$ with probability one. It was shown in \cite{Chapon:14} that the model $\bB_{\text{tar}}+\bV$ corresponds to the fixed rank perturbation models in \ac{RMT} and the spectrum of the sample covariance matrix $\frac{1}{T_s}(\bB_{\text{tar}}+\bV)(\bB_{\text{tar}}+\bV)^{\sf H}$, still converges to a compactly supported probability measure, with up to $K$ outliers that might ``detach'' of the main ``bulk'' of eigenvalues, depending on the system parameters, the \ac{SNR}, the pathgains, and the noise correlation properties \cite{Chapon:14}, \cite{Vinogradova:13}. Hence, the number of such outliers corresponds to the number of targets that can be detected.

\subsubsection{Clutter-plus-noise case}

The total number of clutter clusters $L$ and the number of clutter points $J_l$ within each cluster depend on the system parameters and the propagation environment.
When $L$ and $J_l$ can be considered small as compared to the number of receiver antennas, the targets and the clutter form a subspace of dimension $K+K_{\text{cl}}$ with probability one, small as compared to the number of receiver antennas $N$. The total number of outliers in this case corresponds to the number of detectable targets in addition to the total number of single point clutters. However, small number of clutter points is not a realistic assumption  in general in millimeter wave systems where the clutter tends to be composed of several clusters with a high number of scatter points.

We assume in the following that $J_l$ is of the same order of magnitude as $N$,  for $l=1, \ldots, L$. In this case, the matrix corresponding to the clutter $\bB_{\text{cl}}$, defined in Equation~\eqref{B_cl}, is of full rank equal to $\text{min}(N,T_s,K_{\text{cl}})$ with probability one. We assume that the spectrum of the clutter sample covariance matrix $\frac{1}{T_s}\bB_{\text{cl}}\bB_{\text{cl}}^{\sf H}$ converges to a spectral measure with a compact support composed of up to $L$ intervals, with $L$ fixed as $N, T_s \to \infty$. Using the results from \cite{Dozier:07}, \cite{Marcenko:67}, intuitively, the spectrum of the clutter-plus-noise covariance matrix $\frac{1}{T_s}(\bB_{\text{cl}}+\bV)(\bB_{\text{cl}}+\bV)^{\sf H}$ converges to a limit distribution function with a support composed of compact intervals. The sample eigenvalues forms up to $L+1$ ``bulks'' of eigenvalues corresponding to the clutter-plus-noise part, from which up to $L$ ``bulks'' are due to the clutter. This observation leads to a conjecture that we consider as an assumption in this work. The assumption of large $J_l$ corresponds to a more realistic propagation scenario. Therefore, we summarize the discussed observations into two additional assumptions:
\begin{assumption}
As $J_l\to \infty$, $N \to \infty$, $J_l/N \to \tilde{c}_l>0$, for $l=1, \ldots, L$.
\end{assumption}
\begin{assumption}
As $N \to \infty$, $L$ is fixed and the spectral measure of $\frac{1}{T_s}\bB_{\text{cl}}\bB_{\text{cl}}^{\sf H}$ converges to a probability measure with a compact support composed from up to $L$ intervals.
\end{assumption}

We finally define the sensing \ac{SCNR} defined as 
\begin{equation} \label{SCNR}
\text{{SCNR}} \triangleq \frac{P_s}{\sigma^2+P_{\text{max, cl}}}
\end{equation}
where $P_{\text{max, cl}}$ is largest observed power due to the clutter $P_{\text{max, cl}}$ and $P_s$ is the sensing transmitted power.

\subsection{Hypothesis testing}
We consider the following binary hypothesis test
\begin{equation*}
    \bY = \left\{
    \begin{array}{ll}
\bB_{\text{cl}}+\bV & {\cal{H}}_0 \\
\bB_{\text{tar}} + \bB_{\text{cl}}+\bV & {\cal{H}}_1
    \end{array}
\right.
\end{equation*}
for which under hypothesis ${\cal{H}}_0$ no target is present and under hypothesis ${\cal{H}}_1$ $K$ targets are present. 
As it was observed in \cite{Vinogradova:13} for an transmitting object to be detected, its corresponding transmit power needs to satisfy the so-called detectability limit condition. Similarly, in the case of passive object detection, the TX-s transmit power has to be larger than a limit value, denoted by $P_{\text{lim}}$, which depends on the system dimension parameters, maximum power due to the clutter, the noise variance, and the value of the maximum of the correlation coefficients defined in Equation~\eqref{r_i} if the noise presents temporal correlations.

\subsubsection*{Ratio-based detection test}
Let Assumptions~1-6 hold.
Let $\lambda_1 > \ldots >\lambda_N$ be the eigenvalues of ${\bR}_{T_s}$ ordered in a decreasing order.
Let $K_{\text{max}}<N$ be the upperbound on the number of targets. 
We assume that $P_s>P_{\text{lim}}$.
The estimate of the number of targets is given by
\begin{equation}\label{K_est}
\widehat{K}= \text{arg}~{\text{max}}_{n=1, \ldots, K_{\text{max}}}  \frac{\lambda_n}{\lambda_{n+1}} > 1 + \varepsilon
\end{equation}
where $\varepsilon$ is a threshold depending on the probability of false alarm which is the probability of detecting a target when only clutter-plus-noise is present.

The detection test given by Equation~\eqref{K_est}, based on the ratio of consecutive ordered eigenvalues of the sample covariance matrix ${\bR}_{T_s}$, is similar to the one proposed in \cite{Vinogradova:13} in which a clutter free scenario is considered in the presence of a temporally correlated noise. It was shown in \cite{Baik:05} that the largest eigenvalue of the sample covariance matrix of the correlated noise converges to the Tracy--Widom \cite{Tracy:96} distribution as dimensions grow large. Hence, it is possible, in clutter-free case, to calculate the explicit value of the threshold $\varepsilon$ for a given probability of false alarm. The presence of the clutter further complicates the model and therefore we propose to use an empirical threshold which will depend on a given a probability of false alarm.

\section{Simulation results\label{simu}}
In this section, we present simulation results for target detection using the proposed transmit beamforming schemes. We assume that the number of transmit (sensing and communication) antennas is equal to $M=8$, the number of sensing receiver antennas is $N=16$ (unless it is explicitly mentioned in the figure description or the label), the total number of sensing and communication slots is $T=64$. We analyse the results in terms \ac{ROC} curves, for which correct detection rates versus false alarm rates are plotted. The correct detection rate corresponds to correctly detecting a target under hypothesis ${\cal{H}}_1$ ($K$ targets are present). The false alarm corresponds to the probability of detecting a target under hypothesis ${\cal{H}}_0$ (no target is present). In the plots representing the correct detection rates, the detection threshold is set to a value, for which the probability of false alarm is equal to 0.01. The detection test presented in Equation~\eqref{K_est} is used.

\begin{figure}[h]
	\centering 
	\includegraphics[width=8cm]{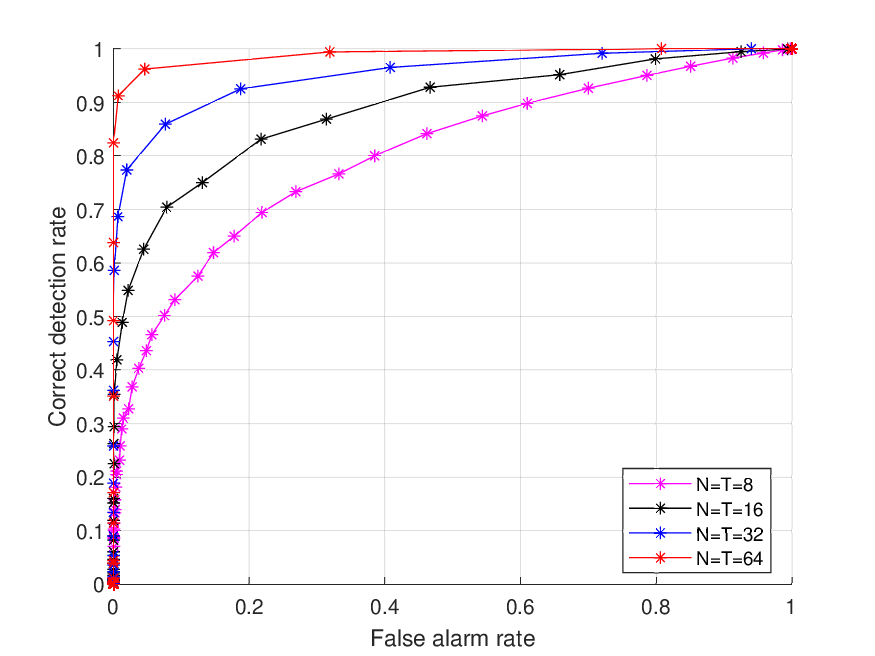}
	\caption{ROC detection curves for different number of sensing receiver antennas for white noise without clutter and $\mathrm{SNR}=-6~\mathrm{dB}$. The number of sensing receiver antennas is considered equal to the number of sensing slots.}
	\label{fig:nrRx}
\end{figure}

\begin{figure}[h]
    \centering 
	\includegraphics[width=8cm]{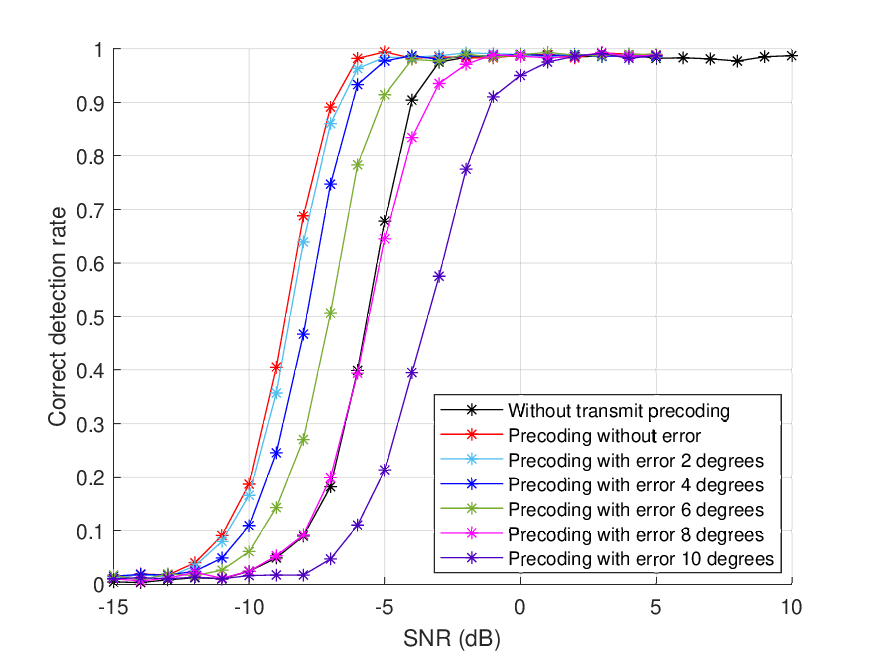}
	\caption{Detection rates versus \ac{SNR} (dB) without and with transmit beamforming for different transmit beamforming angular errors.}
	\label{fig:precoding}
\end{figure}

We consider first a scenario with a single target in the presence of a white noise without a clutter. We assume a beamforming scheme, for which all the power and time slots are allocated fully for sensing, for which both proposed beamforming schemes in Section~\ref{tx_beamforming} are equivalent. In Figure~\ref{fig:nrRx}, the \ac{ROC} curves are plotted for different number of sensing receiver antennas RX-s $N$ for $\mathrm{SNR}=-6~\mathrm{dB}$. $N$ is considered to be equal to the number of sensing time slots, meaning that $N=T_s=T$. We can observe that even for relatively small and realistic values of $N$, the detection of  the target is successful, with a better performance for higher values. 
In Figure~\ref{fig:precoding}, the correct detection rates versus \ac{SNR} (dB) are compared for the transmission without any transmit beamforming applied (in black) and with a beamforming pointing in the direction of the target with a given angular error in degrees. We observe that the detection performance, as expected, is better with a transmit beamforming under condition that the beamforming error does not exceed some limit which is equal to 8 degrees for this scenario (the corresponding curve in magenta).

\begin{figure}[h]
	\centering 
	\includegraphics[width=8cm]{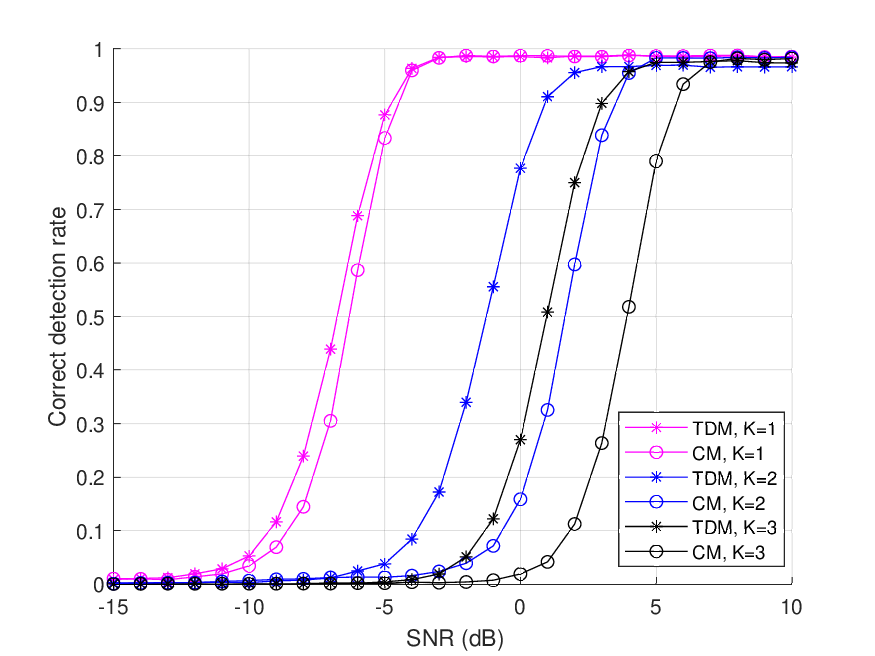}
	\caption{Detection rates for the two proposed beamforming schemes for white noise without clutter versus \ac{SNR} (dB) for different number of targets.}
	\label{fig:K}
\end{figure}

\begin{figure}[h]
	\centering 
	\includegraphics[width=8cm]{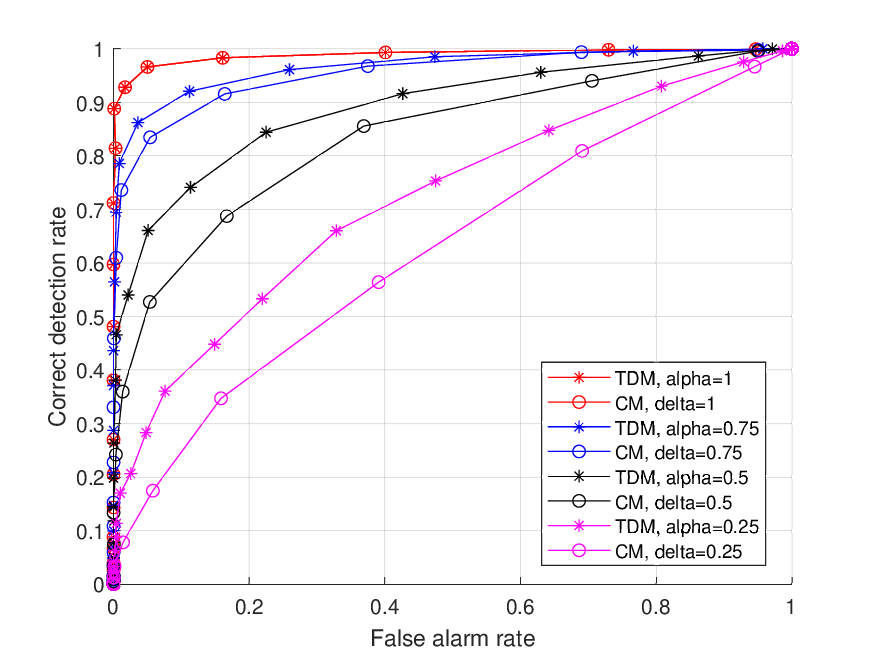}
	\caption{ROC detection curves for different trade-off parameters and the proposed beamforming schemes for white noise without clutter and $\mathrm{SNR}=-6~\mathrm{dB}$.}
	\label{fig:trade_off}
\end{figure}

We compare now the proposed beamforming schemes, denoted by \ac{TDM} and \ac{CM} as introduced in Section~\ref{tx_beamforming} in the presence of a white noise without a clutter. As discussed in Section~\ref{tx_beamforming}, we consider the \ac{JCAS} trade-off parameters $\alpha$ and $\delta$ allowing to set the resource sharing in time and power, respectively, between communication and sensing. In Figure~\ref{fig:K}, we compare the detection rates of the proposed beamforming schemes versus \ac{SNR} (dB) for different number of targets. We assume in this figure that $\alpha=\delta=0.5$, meaning that half of the time, the resources are used for sensing for the \ac{TDM} scheme and half of the power resources is used for sensing for the \ac{CM} scheme. We observe that, in general, the \ac{TDM} scheme outperforms the \ac{CM} beamforming, in particular for higher number of targets. In Figure~\ref{fig:trade_off}, the \ac{ROC} curves for the two schemes are plotted for different values of trade-off parameters for $\mathrm{SNR}=-6~\mathrm{dB}$ and $K=1$. It shows that a higher fraction of allocated resources to sensing leads to smaller differences in performance between the two schemes. For a lower amount of resources available for sensing, the performance of the \ac{TDM} scheme is better than that of the \ac{CM} scheme.
 
\begin{figure}[h]
	\centering 
	\includegraphics[width=8cm]{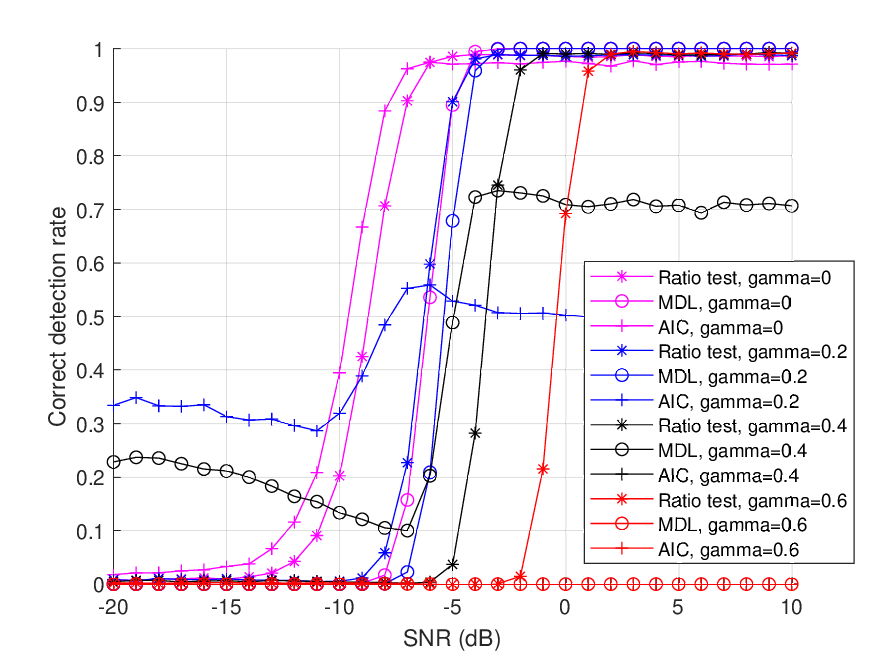}
	\caption{Correct detection rates for different correlation noise parameters versus SNR (dB) for different detection methods.}
	\label{fig:noise}
\end{figure}

We now assume the clutter-free scenario with temporally correlated noise following an autoregressive process of order~1 with the correlation parameter $\gamma$. In Figure~\ref{fig:noise}, the correct detection rates versus \ac{SNR} (dB) are compared for different values of the noise correlation parameter with $K=1$. The proposed ratio-based detector provided in Equation~\eqref{K_est} is compared to the existing \ac{MDL} \cite{Rissanen:78} and \ac{AIC} \cite{Wax:85,Akaike:74} source number estimation methods. In Figure~\ref{fig:noise}, we observe that for the white noise case, the \ac{AIC} yields the best results. However, for the correlated noise case, especially with higher correlation parameter, the \ac{MDL} and \ac{AIC} fail to correctly detect the target and the proposed ratio test scheme is still able to achieve correct detection rates. This demonstrates that using the ratio test is particularly useful in highly correlated noise scenarios.

\begin{figure}[h]
	\centering 
	\includegraphics[width=8cm]{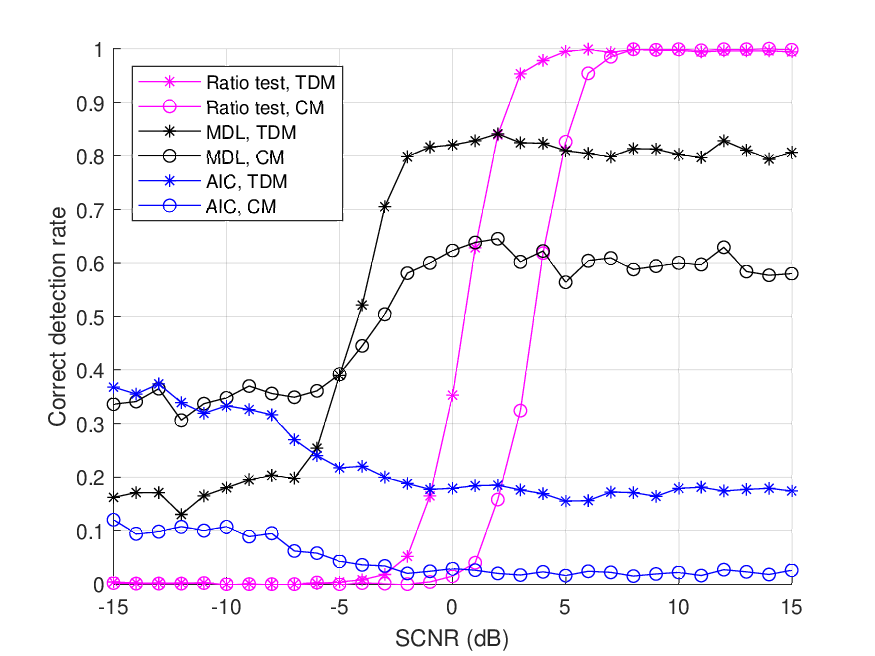}
	\caption{Correct detection rates for the proposed beamforming schemes  versus SCNR (dB) for different detection methods in clutter presence.}
	\label{fig:clutter_scnr}
\end{figure}

\begin{figure}[h]
	\centering 
	\includegraphics[width=8cm]{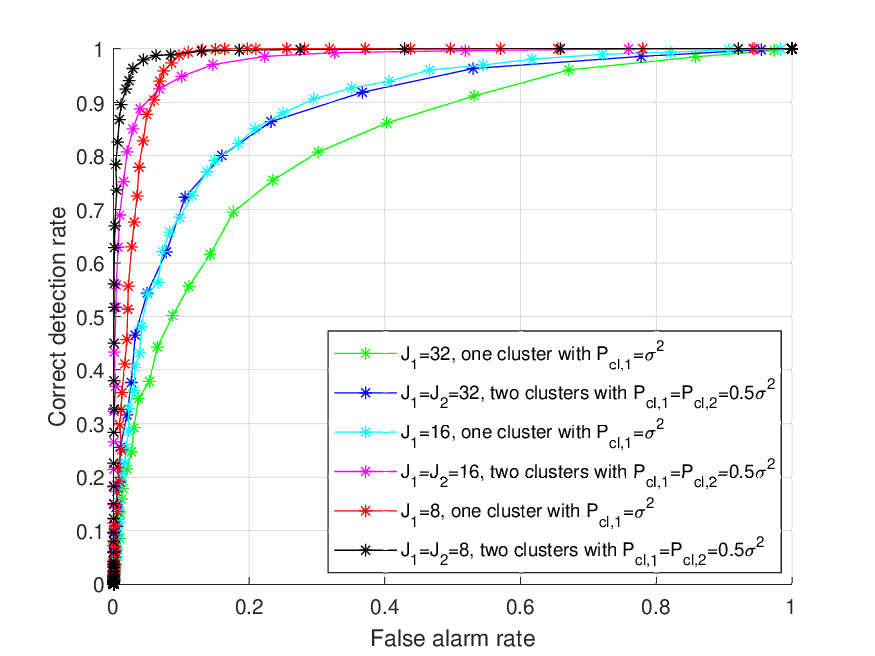}
	\caption{ROC detection curves for different number of clutter clusters and clutter points for $\mathrm{SCNR}=-5~\mathrm{dB}$.}
	\label{fig:clutter}
\end{figure}

We consider now a presence of a clutter composed from two distinct clusters with equal power such that the total power is equal to the noise variance. The number of points for each cluster is equal to $J_1=J_2=32$ with \ac{AoD} and \ac{AoA} spaced with 2 degrees for each single scatter point and the clutter clusters' \ac{AoD} and \ac{AoA} are not overlapping. 
In Figure~\ref{fig:clutter_scnr}, we compare the \ac{TDM} to the \ac{CM} schemes for different detection test methods for $K=1$ and $\alpha=\delta=0.5$. Again, the proposed ratio test is able to correctly detect the target while the \ac{MDL} and \ac{AIC} fail. The \ac{TDM} still gives better performance as compared to the \ac{CM}. In Figure~\ref{fig:clutter}, the presence of the number of clutter clusters is analysed for one cluster having the total clutter power $P_{\text{cl},1}=\sigma^2$ and for two clusters, each with power equal to a half of the noise variance with $P_{\text{cl},1}=P_{\text{cl},2}=0.5\sigma^2$. The equivalent beamforming scheme is assumed for which $\alpha=\delta=1$. We observe that in this scenario, the presence of a strong single clutter degrades the performance more as compared to the presence of two weaker clusters, especially when the number of cluster points gets larger. However, we note that for smaller number of clutter points, the false alarm rate becomes larger. This is due to the fact that for smaller number of clutter points, the Assumption~5 does not hold as the clutter matrix $\bB_{\text{cl}}$ becomes of a fixed rank and the clutter points are detected (wrongly associated to targets) more often under hypothesis ${\cal{H}}_0$.

\section{\label{concl}Conclusions}
\label{Sec:Conc}
We proposed a detection scheme that allows to estimate correctly the number of targets using \ac{JCAS} framework-specific beamforming schemes, especially focusing on beamforming schemes designed for the previously proposed \ac{TDM} and \ac{CM} bistatic \ac{JCAS} modes.
The proposed method was analyzed in the presence of a temporally correlated noise and, eventually, a clutter composed by a number of clusters presenting different angular spreads. The proposed scheme estimates correctly the number of targets whereas the existing schemes, such as the \ac{MDL} and \ac{AIC} fail, especially in the presence of a strong clutter and a heavily correlated noise.
We showed that in general, the \ac{TDM} scheme gives better detection performance compared to the \ac{CM} scheme.

The angular and delay spread phenomenon resulting from the presence of clutter clusters is a crucial problem for target position estimation. Therefore, as a future work, the estimation of the angular and delay spread needs to be addressed to accurately estimate the targets' positions. 

\bibliographystyle{IEEEtran}
\bibliography{JCASv8,FDv3}
\end{document}